\documentclass[12pt]{article}
\input epsf
\usepackage{epsfig}
\title{PT-symmetric eigenvalues for homogeneous potentials}
\author{Alexandre Eremenko\thanks{Both authors are
supported by NSF grant DMS--1665115.}$\;$
and Andrei Gabrielov}
\def\C{\mathbf{C}}
\begin{document}
\maketitle
\begin{abstract} We consider one-dimensional Schr\"odinger equations
with potential $x^{2M}(ix)^\varepsilon,$ where $M\geq 1$ is an integer,
and $\varepsilon$ is real,
under appropriate PT-symmetric boundary conditions.
We prove the phenomenon which was discovered by Bender and Boettcher by
numerical computation: as $\varepsilon$ changes,
the real spectrum suddenly becomes
non-real in the sense that all but finitely many eigenvalues
become non-real. We find the limit arguments
of these
non-real eigenvalues $E$ as $E\to\infty$.

MSC: 81Q05, 34M60, 34A05.

Keywords: one-dimensional Schr\"odinger operators, PT-symmetry,
WKB, asymptotic expansions, entire functions.
\end{abstract}

\noindent
{\bf 1. Introduction.}
\vspace{.1in}

Anharmonic oscillators (one-dimensional Schr\"odinger operators
with polynomial potentials) played an important role in quantum
mechanics since its inception \cite{Heis}, \cite{BJ},
\cite{BW}, \cite{Voros}.
In the early 1990s, Bessis and Zin-Justin made a surprising
conjecture that the eigenvalue problem
$$-w''+ix^3 w=E w,\quad w(\pm\infty)=0,$$
has real discrete spectrum. Their original work is not published
but the history and motivation are described in the survey \cite{DDT3}.
Stimulated by this conjecture, Bender and Boettcher \cite{BB0}, \cite{BB}
proposed to study the one-parametric families of
equations with homogeneous potentials
which contain Airy's equation, harmonic oscillators, and
anharmonic oscillators with cubic and quartic
homogeneous potentials.
This work of Bender and Boettcher started a new large area of
research which is called PT-symmetric quantum mechanics.
PT stands for ``parity and time'', and in our context this means
that the potential $V$
of the operator
\begin{equation}\label{oper}
 -\frac{d^2}{dx^2}+V(x)
\end{equation}
satisfies $V(-\overline{x})=\overline{V(x)}$
and the boundary conditions are interchanged by the transformation
$x\mapsto-\overline{x}$. 

Bender and Boettcher considered the equation
\begin{equation}\label{bb}
-w''+x^{2M}(ix)^\varepsilon w=E w,
\end{equation}
with a positive integer $M$ and real $\varepsilon$. Here the principal branch
of the power is used, so the branch cut in (\ref{bb})
is the positive imaginary ray.
The boundary condition for $\varepsilon=0$ is imposed on
the real line: $w(\pm\infty)=0$, so that for $\varepsilon=0$
we obtain the harmonic oscillator when $M=1$, and a quartic
oscillator when $M=2$.
When $|\varepsilon|$ increases and becomes large,
the boundary condition has to be continuously
deformed (to preserve the discrete spectrum), so that
$w(z)\to 0$ on two {\em normalization rays}
in the complex plane which lay inside
non-adjacent Stokes sectors symmetric with respect to
the imaginary line.

We recall that Stokes sectors are the sectors between
Stokes lines, and Stokes lines
are defined by $V(z)dz^2<0$, where $V$ is the potential in (\ref{oper}).
In the case of a homogeneous potential in (\ref{bb}), Stokes lines are 
equally spaced rays with angles $2\pi/(m+2)$ between adjacent rays,
where
\begin{equation}\label{m}
m=2M+\epsilon.
\end{equation}
The negative imaginary ray is a Stokes ray when $M$ is even,
and bisects a Stokes sector when $M$ is odd.
Rotation of a normalization ray
within its Stokes sector does not change the normalization.
So the normalization conditions just tell us in which pair of
Stokes sectors
eigenfunctions tend to zero.

It is convenient to define the {\em level} of the PT-symmetric
problem as the number of full Stokes sectors between the normalization
rays. Thus the level equals  $M\geq 1$ in (\ref{bb}) for the problem
considered by Bender and Boettcher.

Such eigenvalue problems, with normalization on two rays in the complex
plane, were studied for the first time by Sibuya \cite{Sibuya} for complex
polynomial
potentials,
without any symmetry conditions.

\begin{figure}
\centering
\includegraphics[width=3.0in]{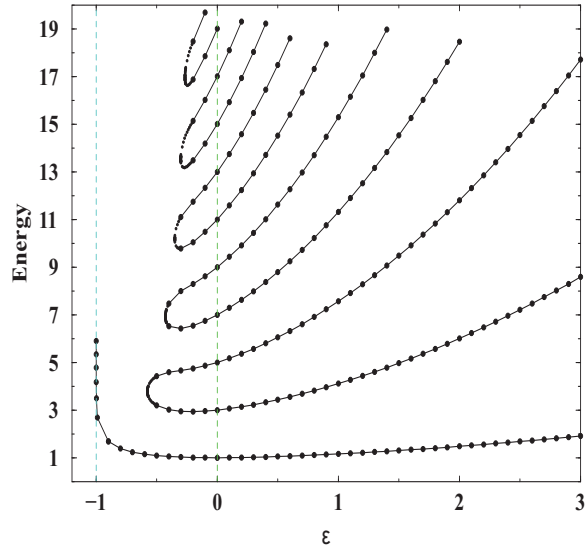}
\caption{Real eigenvalues of $-w''+x^2(ix)^\varepsilon w$ as functions of
$\varepsilon$ (\cite[Fig. 11]{BB}).}\label{fig1}
\end{figure}

\begin{figure}
\centering
\includegraphics[width=3.0in]{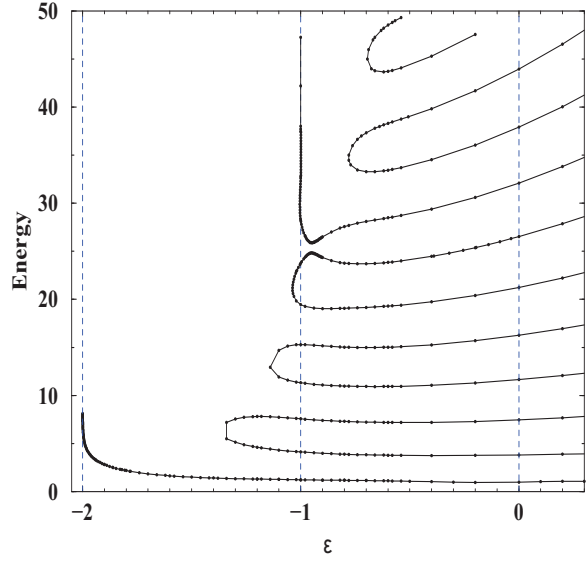}
\caption{Real eigenvalues of $-w''+x^4(ix)^\varepsilon w$ as functions of
$\varepsilon$. The almost vertical part over $\varepsilon=-1$ indicates that
all eigenvalues are real: it actually wiggles about the vertical line
$\varepsilon=-1$ (\cite[Fig. 14]{BB}). }\label{fig2}
\end{figure}

\begin{figure}
\centering
\includegraphics[width=3.0in]{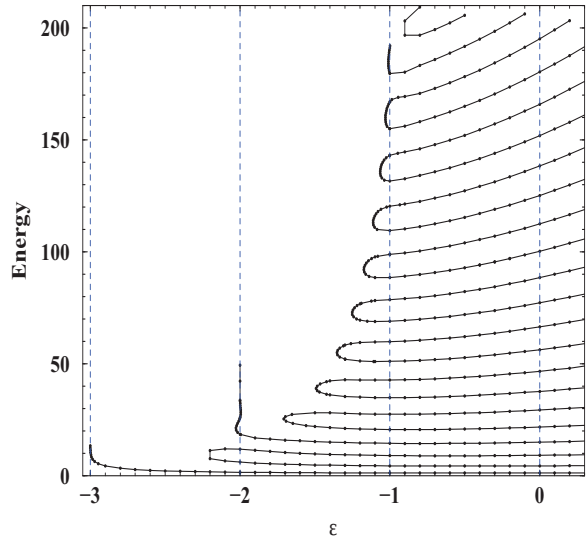}
\caption{Real eigenvalues of $-w''+x^6(ix)^\varepsilon w$ as
functions of $\varepsilon$. All eigenvalues are real for $\varepsilon=-1$
and $\varepsilon=-2$. For non-integer $\varepsilon<0$ almost all
of them
are non-real (\cite[Fig. 20]{BB}).} \label{fig3}
\end{figure}

Eigenvalues of any PT symmetric problem are symmetric with respect
to the real line.
Bender and Boettcher were interested in reality of eigenvalues,
and they obtained by numerical computation
the following remarkable figures: Fig.~\ref{fig1} (level 1), Fig.~\ref{fig2} (level 2)
and Fig.~\ref{fig3} (level 3) which we reproduced from \cite{BB}.

The most conspicuous feature of these figures is a
``phase transition'': for $\varepsilon>0$ all eigenvalues are real, while for
$\varepsilon<0$ they are all real when $\varepsilon$ is an integer
(such that $\varepsilon+2M>1$),
while almost all of them are non-real for negative non-integer
$\varepsilon.$ Here and in what follows
``almost all'' means ``all but finitely many''.

The purpose of this paper is to prove these facts. 
We do this
for $M=1$ and $M=2$, but it seems that our method can be
extended to arbitrary integer $M\geq 1.$

Now we state some known rigorous results about reality of eigenvalues
of (\ref{bb}). In \cite{shin3}, K. Shin proved that that for
any potential $x^{2M}(ix)^\varepsilon$ with integers $M,\varepsilon,\;
2M+\varepsilon>1$,
and for arbitrary PT-symmetric
boundary conditions at infinity, all eigenvalues are real.

Shin's proof consists of two parts: $\varepsilon\geq0$ and $\varepsilon<0$.
The part about $\varepsilon\geq0$ works also for non-integer $\varepsilon$,
see also
\cite{E} for $M=1$ and $M=2$. In the case $\varepsilon<0$, Shin makes
the change of the variable $x\mapsto -x$ to reduce
to the case $\epsilon>0$, which of course can be done
only when $\varepsilon$ is an integer.
In fact, the figures of Bender and Boettcher
and our theorems stated below show that for negative non-integer
$\epsilon$ almost all
eigenvalues are non-real.

All reality proofs are based on the method
invented by Dorey, Dunning and Tateo, who proved the
original conjecture of Bessis and Zinn-Justin in \cite{DDT0}.
These authors discovered that spectral determinants (entire functions
whose zeros are eigenvalues) arising in
PT-symmetric eigenvalue problems that we consider here
arise also in integrable models
of
statistical mechanics on 2-dimensional lattices.
This gives an additional reason to study these entire functions.
A comprehensive
survey of their
results on this ODE/IM correspondence is given in \cite{DDT}.

The case of non-integer $\varepsilon<0$ was considered in
\cite{DDT2}, but in that paper the authors only study the first few
eigenvalues, and small values of $\varepsilon.$

Our results are the following.
\vspace{.1in}

\noindent
{\bf Theorem 1.} {\em For the equation (\ref{bb}) with $M=1$,
$\varepsilon\in (-1,0)$ and the boundary conditions
$w(x)\to 0,\; x\to\pm\infty$, almost all eigenvalues are non-real.
The arguments of the eigenvalues accumulate to
$$\pm\pi\frac{2-m}{2+m},\quad\mbox{as}\quad E\to\infty,$$
where $m=2M+\epsilon$.
}
\vspace{.1in}

\noindent
{\bf Theorem 2.} {\em For the equation (\ref{bb}) with $M=2$,
$\varepsilon\in(-3,0)\backslash \{-1,-2\}$ and the boundary conditions
\begin{equation}\label{bc}
w(x)\to 0,\quad x\to\infty,\quad
\arg x=-\frac{\pi}{2}\pm\frac{3\pi}{m+2},
\end{equation}
almost all eigenvalues
are non-real, and their arguments accumulate to
$$\pm\pi\frac{4-m}{2+m},\quad\mbox{as}\quad E\to\infty,$$
where $m=2M+\epsilon$.}
\vspace{.1in}

When $\epsilon\in(-2,0)$ the boundary condition (\ref{bc}) is equivalent to
$w(x)\to 0,\; x\to\pm\infty$.

Our method is different from that of \cite{DDT2}.
The plan is the following. First we introduce an auxiliary
self-adjoint problem
with real discrete spectrum, and an entire function $f$ which is called
a {\em spectral determinant} whose zeros are exactly the eigenvalues
of this self-adjoint problem. Improving the previous results of
Sibuya and Shin, we obtain an asymptotic expansion of this spectral
determinant $f$. Then we use the important
fact that the spectral
determinants of the PT-symmetric problems are also
entire functions and that they
have explicit expressions in terms of
$f$, see \cite{Tabara,DDT,shin3,E}.
By studying the asymptotic behavior of these spectral determinants
we can make conclusions about the asymptotic distribution
of their zeros as $E\to\infty$. This permits us to find their limit
arguments. To show that there are only finitely many real eigenvalues,
we combine the asymptotics with the ``cancellation conditions''
in the formulas for the spectral determinants.

In \cite{BB} and \cite{DDT2} doubts are expressed about applicability of
the WKB method for non-integer exponent (\ref{m}), especially
when $m<2$. In fact the proof
of our main technical result, Theorem 3, can be considered as a
refinement of the usual WKB method, adapted for handling non-integer $m$.

When $M=1,\;\varepsilon\to -1$ and when $M=2,\;\varepsilon\to -2$,
the Stokes sectors containing the normalization rays became adjacent
and the limit problem (for the Airy equation in the first case and
for the quadratic potential in the second) cannot have eigenvalues,
so all eigenvalues escape to infinity.

Numerical evidence of what happens when $\varepsilon$ is below these lower
limits is presented in the recent paper \cite{Bn}.
The method of our paper
breaks down when $m=2M+\varepsilon\leq 1$.
\vspace{.1in}

\noindent
{\bf 2. Auxiliary self-adjoint problem.}
\vspace{.1in}

Consider the differential equation
\begin{equation}\label{1}
y''=(z^m+\lambda)y,\quad\mbox{where}\quad m>1,\; m\neq 2.
\end{equation}
All powers and logarithms are understood in
the sense of principal branches: $|\arg z|<\pi$, unless
some other branch is explicitly specified.
\vspace{.1in}

\noindent
{\bf Sibuya's theorem.}
{\em For every $\lambda\in\C$ there exists a unique solution
$y_0(z,\lambda)$ on the positive ray such that
\begin{equation}\label{2}
y_0(z,\lambda)=(1+o(1))z^{-m/4}\exp\left(-\frac{2}{m+2}z^{(m+2)/2}-
\frac{\lambda}{2-m}z^{(2-m)/2}\right),
\quad z\to+\infty.
\end{equation}
This solution has an analytic continuation in $z$
to $\C\backslash(-\infty,0]$,
and the asymptotic formula (\ref{2}) holds in every sector
$$\{ z:|\arg z|\leq 3\pi/(m+2)-\epsilon\},\quad\epsilon>0.$$
When $m>2$, the second term under the exponent in (\ref{2}) is redundant.

The limits
$$f(\lambda):=\lim_{z\to 0+}y_0(z,\lambda),\quad\mbox{and}\quad
f_1(\lambda):=\lim_{z\to 0+}y^\prime_0(z,\lambda)$$
exist, and both are
entire functions of $\lambda$, of order
\begin{equation}\label{rho}
\rho=\frac{1}{2}+\frac{1}{m}.
\end{equation}
}
\vspace{.1in}

\noindent{\em Comments.}
Sibuya stated this theorem for integer $m$, but his proof
in \cite[Ch.~2]{Sibuya} requires only
minor modification to deal with real $m>1$. In
\cite{DDT,DDT2,Tabara}, Sibuya's theorem is used for non-integer $m$.
We now explain the necessary modifications.

As the first step, one changes the
independent variable to $\zeta=\sqrt{z}$
and by some changes of the dependent variables reduces (\ref{1}) to
the following Riccati equation:
\begin{eqnarray*}
\zeta^{-m-1}p'(\zeta)&=&(\lambda\zeta^{-2m}+(m/2)\zeta^{-2-m})p^2
+(4+2\lambda\zeta^{-2m})p\\
&+&\lambda\zeta^{-2m}-(m/2)\zeta^{-2-m}.
\end{eqnarray*}
Then one shows \cite[Lemma 9.1]{Sibuya}
that this Riccati equation has a formal solution
of the form
$$p(\zeta)=\sum_{n=1}^\infty p_n\zeta^{-n}.$$
This is where the difference between integer and non-integer $m$ arises.
When $m$ is not an integer, one considers the additive semi-group
\def\Z{\mathbf{Z}}
\begin{equation}\label{Lambda}
\Lambda=\{ k_1+k_2m: k_1,k_2\in\Z_{\geq 0}\}
\end{equation}
and
finds a unique formal solution of the Riccati equation in the form
\begin{equation}\label{beta}
p(\zeta)=\sum_{\beta\in\Lambda} p_\beta\zeta^{-\beta}.
\end{equation}
It is easy to see that such a formal solution exists and is unique.
Then one checks that the basic results on the asymptotic expansions
(theorems 3.3, 4.1 and 5.1 in \cite{Sibuya}) extend without changes
to the asymptotic expansions of the form (\ref{beta}).
For such general asymptotic expansions, see, for example,
\cite{Bourbaki}.

The concluding step, showing that there is a true solution corresponding
to the asymptotic expansion (\ref{beta}), and this solution is an entire
function of $\lambda$, is achieved by reducing the Riccati equation
to an integral equation and solving this integral equation by
iteration. This last step requires no changes for non-integer $m$.
\vspace{.1in}

Sibuya \cite[Thm. 19.1]{Sibuya} also studied the
asymptotics of $f$ and $f_1$,
and found the principal terms
$$f(\lambda)=(1+o(1))\lambda^{-1/4}\exp\left( K_m\lambda^\rho\right),$$
$$f_1(\lambda)=-(1+o(1))\lambda^{1/4}\exp\left( K_m\lambda^\rho\right),$$
for $m>2$. Here $K_m>0$ is an explicit constant.
The error term estimates were improved by Shin \cite{shin1} who proved
\begin{equation}\label{sh1}
f(\lambda)=(1+O(\lambda^{-\rho}))
\lambda^{-1/4}\exp\left( K_m\lambda^\rho\right),
\end{equation}
\begin{equation}\label{sh2}
f_1(\lambda)=-(1+O(\lambda^{-\rho}))\lambda^{1/4}
\exp\left( K_m\lambda^\rho\right).
\end{equation}
These asymptotics hold when $|\lambda|\to\infty$,
uniformly in every sector
\begin{equation}\label{sector}
\{\lambda:|\lambda|\leq\pi-\epsilon\}.
\end{equation}
They can be also extended to a sector containing the negative ray in the following way:
\begin{eqnarray}\label{shneg}
f(\lambda)&=&\left(1+O(\lambda^{-\rho})\right)
\lambda^{-1/4}\exp\left(K_m\lambda^\rho\right)\\ \label{shneg1}
&+&%
\left(i+O(\lambda^{-\rho})\right)\lambda^{-1/4}
\exp(e^{-2\pi i\rho}K_m \lambda^\rho),
\end{eqnarray}
as $\lambda\to\infty,\; \pi-\epsilon<\arg\lambda< \pi+\epsilon$,
with some $\epsilon>0$.
Similar result holds for $f_1$.
Asymptotics near the negative ray permitted Shin
\cite[Cor. 2.1]{shin1} to locate
the zeros of $f$ with the error smaller than the
distance to the nearest other zero. These zeros are
negative (see below).
All these results of Sibuya and Shin generalize without
changes to the case of non-integer $m>1$, except that $K_m<0$
for $m\in(1,2)$.

We will need the following more precise asymptotics of $f$.
\vspace{.1in}

\noindent
{\bf Theorem 3.} {\em Suppose that $m>1$, and $m$ is not an integer.
Function $f$ in Sibuya's theorem has all
zeros negative, and the
following asymptotics holds
\begin{equation}\label{3a}
f(\lambda)=\lambda^{-1/4}\Psi(\lambda^\rho)\exp\left( K_m\lambda^\rho\right),
\end{equation}
where
\begin{equation}
\label{4a}\Psi(\mu)=\left(1+\sum_{n=1}^{[m]}c_n\mu^{-n}+(1/8)
\Gamma(m+1)\mu^{-m}+O(\mu^{-\kappa})\right),
\end{equation}
as $\mu=\lambda^\rho$, and $|\lambda|\to\infty$ in every sector (\ref{sector}).
Here $\kappa>m$.}
\vspace{.1in}

Explicit expressions for $K_m,c_1$ will be given later,
they are irrelevant for our main argument, except the sign of $K_m$.
Presence of
the term $\lambda^{-m\rho}$ in the asymptotics is crucial.

\vspace{.1in}

{\em Proof.} First we show that all zeros of $f$ are negative.
Indeed, if $f(\lambda)=0$ then $-\lambda$ is an eigenvalue
of the Sturm-Liouville problem on the positive ray:
$$-y''+z^my=-\lambda y,\quad y(0)=y(+\infty)=0,$$
with positive potential, so $-\lambda>0$.

\begin{figure}
\centering
\includegraphics[width=2.5in]{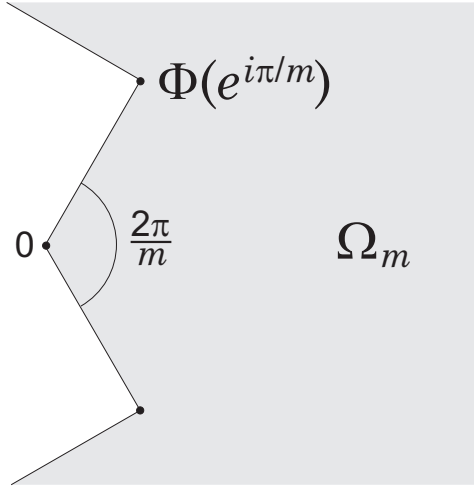}
\caption{The region $\Omega_m$ for $m=3$.}\label{fig:omega}
\end{figure}

In the equation (\ref{1}), we make the change of the variable
\begin{equation}\label{w}
w(z)=y(\lambda^{1/m}z),\quad y(z)=w(\lambda^{-1/m}z),
\end{equation}
and obtain
\begin{equation}\label{mu}
w''=\mu^2(z^m+1)w,\quad\mbox{where}\quad\mu:=\lambda^\rho,
\end{equation}
where $\rho$ is defined in (\ref{rho}).
So we consider the differential equation
\begin{equation}\label{5}
w''=\mu^2Q(z)w,\quad\mbox{where}\quad Q(z)=z^m+1.
\end{equation}
Following Liouville, we introduce the new independent variable
\begin{equation}\label{zeta}
\zeta=\Phi(z)=\int_0^z\sqrt{Q(t)}dt.
\end{equation}
This function is well defined in
the sector $|\arg z|<\pi/m$, and maps this
sector onto the region $\Omega_m$ in the $\zeta$-plane (see Fig.~\ref{fig:omega}), which contains the
positive ray and is symmetric with respect to the real line.
If $m\in (1,2)$ this region $\Omega_m$ overlaps itself,
and has to be considered
as a Riemann surface spread over the $\zeta$-plane.

On the real line $\Phi$ is strictly increasing, $\Phi'(0)=1$ and
$\Phi(+\infty)=+\infty.$
Notice that the region $\Omega_m$ contains the sector $|\arg\zeta|<\pi/m$.

Function $\Phi$ has a convergent expansion near $0$ with exponents
of the set $\Lambda$
as in (\ref{Lambda}):
\begin{equation}\label{Phi-exp}
\Phi(z)=z+\frac{1}{2(m+1)}z^{m+1}+\ldots.
\end{equation}
The asymptotics of $\Phi$ at $\infty$ is
\begin{equation}\label{Phi-infty}
\Phi(z)\sim\frac{1}{m+1}z^{(m+2)/2},\quad z\to\infty.
\end{equation}

We introduce a new function $u(\zeta,\mu)$ in $\Omega_m$:
\begin{equation}\label{u}
u(\zeta,\mu)=Q^{1/4}(z)w(z),\quad\mbox{where}\quad z=\Phi^{-1}(\zeta).
\end{equation}
According to Liouville, see for example \cite{Heading}, this function
satisfies
\begin{equation}\label{6}
u''=\mu^2u+gu,\quad\mbox{where}\quad u''=d^2u/d\zeta^2,
\end{equation}
and
\begin{equation}
\label{7}
g(\zeta)=
\left(-\frac{5}{16}\frac{{Q'}^2}{Q^3}+
\frac{Q''}{4Q^2}\right)\circ\Phi^{-1}(\zeta).
\end{equation}
Notice that $g$ is integrable on every ray in $\Omega_m$,
in fact a simple calculation using (\ref{Phi-infty}) shows that
\begin{equation}\label{z2}
g(\zeta)=O(\zeta^{-2}),\quad\zeta\to\infty,\quad\zeta\in \Omega_m.
\end{equation}
Near the origin, $g$ has a convergent expansion with exponents
in $\Lambda-2$:
\begin{equation}\label{near}
g(\zeta)=\frac{1}{4}m(m-1)\zeta^{m-2}+\ldots.
\end{equation}
Considering $gu$ as the right hand side of the non-homogeneous linear differential
equation (\ref{6}) and using the variation of constants formula, we
obtain an integral equation for $u$:
\begin{equation}\label{8}
u(\zeta,\mu)=e^{-\mu\zeta}+\frac{1}{\mu}\int_\zeta^\infty\sinh\mu(t-\zeta)
g(t)u(t,\mu)dt.
\end{equation}
Here $\zeta\in \Omega_m$ and the integration is along any path in $\Omega_m$
from $\zeta$ to $\infty$.

It is easy to check directly that
a solution of this integral equation, if exists, satisfies (\ref{6}).
We simplify our integral equation by introducing the function
\begin{equation}\label{F}
F(\zeta,\mu)=e^{\mu\zeta}u(\zeta,\mu).
\end{equation}
Then
\begin{equation}\label{9}
F(\zeta,\mu)=1+\frac{1}{2\mu}\int_\zeta^\infty\left(1-e^{\mu(\zeta-t)}\right)g(t)F(t,\mu)dt.
\end{equation}
Suppose first that $\lambda,\mu$ and $\zeta$ are positive,
and the integration in (\ref{9}) is along the positive ray.
We will construct a solution $F$
of this integral equation in $\Omega_m$
which tends to $1$ as $\zeta\to+\infty$.
Then we will set according to our changes of the variables
\begin{equation}\label{wkb}
y_0(z,\lambda)=c(\lambda)Q^{-1/4}(\lambda^{-1/m}z)\exp\left(-\lambda^\rho
\Phi(\lambda^{-1/m}z)\right)F(\Phi^{-1}(\lambda^{-1/m}z),\lambda^\rho),
\end{equation}
where
\begin{equation}\label{C}
c(\lambda)=\lambda^{-1/4}e^{K_m\lambda^\rho},
\end{equation}
and
$$K_m=\int_0^\infty\left(\sqrt{t^m+1}-t^{m/2}\right)dt>0,
\quad\mbox{when}\quad m>2,$$
$$K_m=\int_0^\infty \left(\sqrt{t^m+1}-t^{m/2}-(1/2)t^{-m/2}\right)dt<0,\quad
\mbox{when}\quad 1<m<2.$$
It is easy to verify using (\ref{F}), (\ref{u}), (\ref{w}) and
(\ref{zeta}), that this solution $y_0$ has the
correct asymptotics (\ref{2}) as $z\to+\infty$.
Expression (\ref{wkb}) is the WKB solution of (\ref{1}),
$$y_0(z,\lambda)=(c_1(\lambda)+o(1))(z^m+\lambda)^{-1/4}\exp\left(
-\int_0^z\sqrt{t^m+\lambda}\, dt\right),$$
where the constant factor $c_1(\lambda)=e^{K_m\lambda^\rho}$
is chosen so that
the dependence of the exponential term in the
asymptotics (\ref{2}) on $\lambda$ is as simple
as possible: when $m>2$ it does not depend on $\lambda$ at all,
while for $m\in(1,2)$ it is a simple entire function of $\lambda$.

It is easy to verify directly that $y_0$ defined by (\ref{wkb})
formally satisfies (\ref{1}) if $F$ satisfies (\ref{9}).
\vspace{.1in}

{\em Existence of a solution of (\ref{9})}.
\def\I{\mathbf{I}}
We set $F_0=0,\; F_n=\I_\mu(F_{n-1}),$ where $\I_\mu$ is the integral operator
in the right hand side of (\ref{9}). We assume first that $\mu>0$, $\zeta>0$ and
that integration is over the positive ray. Since
$$|F_n-F_{n+1}|(\zeta)\leq\frac{1}{2\mu}\int_0^\infty|g(t)||F_n(t)-F_{n-1}(t)|dt\leq\frac{1}{2\mu}\| g\|_1\| F_n-F_{n-1}\|_\infty,$$
the series
$$F=F_1+(F_2-F_1)+(F_3-F_2)+\ldots$$
converges uniformly on $[0,\infty)$ when $\mu>\mu_0$, and defines a bounded
function $F$ solving (\ref{9}). We also have $|F-1|\leq \mathrm{const}/\mu$.
\vspace{.1in}

{\em Asymptotic expansion of $F(0,\mu)$ for $\mu>0.$}
Equation (\ref{9}) is of the form
\def\T{\mathbf{T}}
\def\1{\mathbf{1}}
$$F=\1+\frac{1}{2\mu}\T_\mu[F],$$
where $\T_\mu=2\mu(\I_\mu-\mathbf{1})$ is the integral operator
$$\T_\mu[F](\zeta)=
\int_\zeta^\infty\left(1-e^{\mu(\zeta-t)}\right)g(t)F(t)dt.
$$
This equation is satisfied by the formal series
\begin{equation}\label{formal}
F=\mathbf{1}+\sum_{n=1}^\infty (2\mu)^{-n}\T^n_\mu [\mathbf{1}].
\end{equation}
We set $\zeta=0$ in this series, and obtain
$$\T_\mu[\1](0)=\int_0^\infty\left(1-e^{-\mu t}\right)g(t)dt,$$
\begin{equation}\label{T2}
\T^2_\mu[\1](0)=\int_0^\infty\left(1-e^{-t\mu}\right)g(t)\int_t^\infty
\left(1-e^{\mu(t-t_1)}\right)g(t_1)dt_1dt,
\end{equation}
and so on,
\begin{eqnarray}\label{Tn}
\T^n_\mu[\1](0)&=&\int_0^\infty\left(1-e^{-\mu t}\right)g(t)\int_t^\infty
\left(1-e^{\mu(t-t_1)}\right)g(t_1)\\
\nonumber
&\times&\int_{t_1}^\infty\left(1-e^{\mu(t_1-t_2)}\right)g(t_2)\ldots\\
\nonumber
&\times&\int_{t_{n-2}}^\infty\left(1-e^{\mu(t_{n-2}-t_{n-1})}\right)g(t_{n-1})
dt_{n-1}dt_{n-2}\ldots dt_1dt.
\end{eqnarray}
The $n$-th term of the series (\ref{formal}) is $O(\mu^{-n})$, because
$\| g\|_1<\infty$ and $F$ is bounded. So to obtain an asymptotic expansion
of $F(0,\mu)$ it is sufficient to obtain an asymptotic expansion
of each integral $\T_\mu^n[\1](0)$.

We write
\begin{equation}\label{s}
\T_\mu[\1](0)=\int_0^\infty g(t)dt+\int_0^\infty e^{-\mu t}g(t)dt.
\end{equation}
The first of these integrals is constant:
$$c_1:=\int_0^\infty\left(\frac{5}{16}\frac{m^2x^{2m-2}}{(x^m+1)^3}+
\frac{m(m-1)x^{m-2}}{4(x^m+1)^2}\right)d\zeta=
\frac{m}{32}B(2-1/m,1/2+1/m),$$where $d\zeta=\sqrt{Q(x)}dx$,
and $B$ is Euler's Beta-function.

For the second integral in (\ref{s}) we use the following lemma.
Let us say that a function $h(x)$ expands into a convergent series of real
powers if
\begin{equation}\label{conv}
h(x)=\sum_{n=0}^\infty a_nx^{\beta_n},\quad 0<x<\delta,
\end{equation}
where the exponents $\beta_n\to+\infty$ are real,
and the convergence is absolute and uniform.
\vspace{.1in}

\noindent
{\bf Lemma 1.} {\em Suppose that $h$ is integrable on the real line,
and expands into a convergent series of real
powers as in (\ref{conv}). Then
$$\int_0^\infty e^{-\mu t}h(t)dt$$
has an asymptotic expansion
$$\sum_{n=0}^\infty a_n\Gamma(\beta_n+1)\mu^{-\beta_n-1}.$$}

{\em Proof of Lemma 1.}
$$\int_0^\infty e^{-\mu t}h(t)dt=\int_0^\delta
e^{-\mu t}h(t)dt+O(e^{-\delta\mu})
=\sum_{n=0}^\infty a_n\int_0^\delta e^{-\mu t}t^{\beta_n}dt+
O(e^{-\delta\mu}),$$
and
$$\int_0^\delta e^{-\mu t}t^{\beta_n}dt=\int_0^\infty e^{-\mu t}t^{\beta_n}dt+
O(e^{-\delta\mu})
=\mu^{-\beta_n-1}\Gamma(\beta_n+1)+O(e^{-\delta\mu}).$$
This proves Lemma 1.
\vspace{.1in}

Applying Lemma 1 to the second integral in (\ref{s}) we first
expand $g$ in (\ref{7}) and obtain
$$g(t)=\frac{1}{4}m(m-1)t^{m-2}+O(t^{m-1}).$$
Then Lemma 1 implies that
$$\int_0^\infty e^{-\mu t}g(t)dt=\frac{1}{4}m(m-1)\Gamma(m-1)\mu^{1-m}+O(\mu^{-m}),$$
and this gives
$$F(0,\mu)=1+c_1\mu^{-1}+\frac{1}{8}\Gamma(m+1)\mu^{-m}+
O(\mu^{-m-1})+O(\mu^{-2}).$$
This proves (\ref{4a}) when $m\in(1,2)$.

When $m>2$ one needs to show that for $n\geq 2$
the largest non-integer power of $\mu$
in the expansion of $\mu^{-n}\T^n_\mu[\1](0)$
is less than $-m$. To do this, we make the change of the variable
in (\ref{Tn}):
$$x_1=t,\quad x_2=t_1-t,\quad x_3=t_2-t_1,\quad\ldots\quad
x_n=t_{n-1}-t_{n-2},$$
and break the integral (\ref{Tn}) into the sum of $2^n$ integrals of
the form
\def\x{\mathbf{x}}
\begin{equation}\label{In}
I_{n,J}(\mu)=\int_0^\infty\ldots\int_0^\infty e^{-\mu L(\x)}g(x_1)\ldots
g(x_1+\ldots+x_n)\, dx_n\ldots dx_1,
\end{equation}
where $L(\x)=\sum_{j\in J}x_j,$ and $J$ is a subset of $\{1,\ldots,n\}$.
Then we have the following
\vspace{.1in}

\noindent
{\bf Lemma 2.} {\em If $g$ is an analytic function in the sector
$|\arg z|<\epsilon$ satisfying
\begin{equation}\label{con}
g(z)=O(z^\alpha),\quad z\to 0,\quad\mbox{and}\quad g(z)=O(z^{-2}),\quad z\to\infty
\end{equation}
where $\alpha>0$, then
$$\mu^{-n}I_{n,J}=\sum_{k=0}^{[\alpha]+n+1}c_k\mu^{-k}+O(\mu^{-\alpha-1-n}),
\quad\mu\to\infty.$$}
\vspace{.1in}

Our function $g$ satisfies conditions (\ref{con}) with $\alpha=m-2$,
so the contribution from $\mu^{-n}\T^n_\mu[\1](0)$ to (\ref{4a})
consists of integer powers and powers less than $-m$ of $\mu$,
which will prove (\ref{4a}).
\vspace{.1in}

{\em Proof of Lemma 2.} (F. Nazarov, private communication).
Notice that derivatives $g^{(k)}$ are bounded on $[0,\infty)$ for
$k\leq [\alpha]$. Moreover, all these derivatives satisfy
\begin{equation}\label{der}
g^{(k)}(z)=O(z^{-2}),\quad z\to\infty,
\end{equation}
because $g$ is analytic in a sector and we can apply Cauchy's estimate
to them. Put
\begin{equation}\label{q}
q=[\alpha]-1,
\end{equation}
and write the Taylor formula with remainder for $g$:
\begin{equation}\label{taylor}
g(u+v)=g(u)+g'(u)v+\ldots+g^{(q)}(u)\frac{v^q}{q!}+\epsilon(u,v),
\end{equation}
where
\begin{equation}\label{er}
|\epsilon(u,v)|\leq Mv^{q+1}.
\end{equation}
This formula (\ref{taylor}), (\ref{er}) holds {\em for all} positive $u,v$,
because $g^{(q+1)}$ is bounded on the whole positive ray.

For each $g(x_1+\ldots+x_k)$ in our integral, we split the argument
into two summands:
$$u_k=\sum_{j\not\in J, j\leq k}x_j\quad\mbox{and}\quad
v_k=\sum_{j\in J,j\leq k}x_j,$$
and apply (\ref{taylor}) to 
$g(v_k+u_k)$. Notice that for each $k$
\begin{equation}\label{y}
v_k\leq L(\x).
\end{equation}
Now we multiply these expressions obtained from the Taylor formula applied
to each $g(x_1+\ldots+x_k)$, and split the integral into summands.

There are two types of summands.
\vspace{.1in}

\noindent
{\em Type 1.} Those which do not contain any error term $\epsilon$.
They are of the form
$$\int_0^\infty\ldots\int_0^\infty e^{-\mu L(\x)}g^{(j_1)}(u_1)\ldots g^{(j_n)}(u_n)
v_1^{j_1}\ldots v_n^{j_n}d\x.$$
Notice that each variable $x_j$ participates either in some $u_k$ or
in the exponential factor. This is because
in $g(x_1+\ldots+x_n)$ all variables participate, and some
term from the Taylor expansion of this function will enter each
summand. So the integral is convergent in view of (\ref{z2}).
Integration in the variables that are not in $J$ gives a constant factor,
and the integration in the variables which are in $J$ gives integer
power of $\mu$, because $v_j$ are honest polynomials
(with integer powers of these variables).
\vspace{.1in}

\noindent
{\em Type 2.} Those terms which contain at least one $\epsilon$. These terms
are estimated using (\ref{er}) and (\ref{y}).
All previous arguments are
applicable to the {\em estimates}, (the estimates contain only
integer powers), and these terms contribute
to the asymptotics at most
$$\mu^{-q-2}<\mu^{-\alpha-1}$$
by Lemma 1\footnote{We don't really need Lemma 1 here, but the trivial fact
that if $P(x)$ is a monomial of degree $d$ in the variables
participating in $J$, then
$$\int\ldots\int e^{-\mu L(\x)}P(\x)d\x=c\mu^{-d-1}.$$} and (\ref{q}). Finally,
the whole integral is divided by $\mu^n$
and we get
that the top non-integer power in $I_{n,J}$ is at most
$$\mu^{-\alpha-1-n}.$$
This proves Lemma 2.
\vspace{.1in}

{\em Analytic continuation and extension of the asymptotics
to complex $\mu$.}
\vspace{.1in}

First we keep $\mu$ positive and perform an analytic continuation of
$\zeta\mapsto F(\zeta,\mu)$ to the sector $|\arg\zeta|<\pi/m$.
\def\Rea{\mathrm{Re}\, }
To do this we choose the
path of integration in (\ref{9}) to consist of two segments $[\zeta,\zeta_0]$
and $[\zeta_0,+\infty)$, where $\zeta_0>\Rea\zeta.$ This defines our function
$F(\zeta,\mu)$ in the sector, and $F(\zeta,\mu)\to 1$ as $\zeta\to\infty$
in this sector. Next we notice that the contour of integration
can be changed to a ray $\{\zeta+re^{i\phi}:r\geq 0\}$,
as long as $|\phi|<\min\{\pi/m,\pi/2\}$, without affecting the equation.
This follows from Cauchy's theorem.

Once a ray of integration in (\ref{9}) is fixed, the arguments
in the previous part remain valid as long as $|e^{-\mu \zeta}|$
is decreasing on the ray of integration and tends to zero exponentially
as $\zeta\to\infty$.
For example, if the integration is along the positive ray, the asymptotics
is valid for $|\arg\mu|<\pi/2$. By rotating the integration ray, as
explained above, we obtain the asymptotic expansion
$$F(0,\mu)=1+\sum_{n=1}^{[m]}c_n\mu^n+(1/8)\Gamma(m+1)\mu^m+O(\mu^{-m-\epsilon})$$
which is valid in the
sector $|\arg\mu|<\pi/2+\pi/m-\epsilon,$ for any $\epsilon>0$.

Recalling that $\lambda=\mu^{1/\rho}$ we conclude that the
required asymptotic expansion (\ref{4a}) holds in the sector $|\arg\lambda|<\pi-\epsilon.$

This completes the proof of Theorem 1.

\vspace{.2in}

\noindent
{\bf Problem of level 1. Proof of Theorem 1.}
\vspace{.2in}

We consider the equation
\begin{equation}\label{eq1}
y''=(z^m+\lambda)y
\end{equation}
with $1<m<2$ and the boundary conditions
\begin{equation}\label{bc1}
y(re^{\pm2\pi i/(m+2)})\to 0,\quad r>0,\quad r\to\infty.
\end{equation}
This is obtained by the change of the variable $z=ix$ in (\ref{bb}),
and we have $\lambda=-E$, $m=2+\varepsilon$.
The problem does not change when we rotate the normalization rays
within their Stokes sectors, and $m\in(1,2)$, so the boundary condition (\ref{bc1})
is equivalent to $y(\pm ir)\to 0$ as $r>0,\; r\to\infty$.

If $y_0(z,\lambda)$ is a solution of (\ref{eq1}) then
$$y_k(z,\lambda)=\omega^{k/2}y_0(\omega^{-k}z,\omega^{2k}\lambda),$$
where
$$\omega=e^{2\pi i/(m+2)},\quad \omega^{k/2}=e^{\pi i k/(m+2)},$$
is also a solution. We take as $y_0$ the solution normalized by
(\ref{2}).

Problem (\ref{eq1}), (\ref{bc1}) has discrete spectrum,
and an entire function $C(\lambda)$
with zeros at the eigenvalues, which is called the spectral determinant,
can be found from the following equation:
$$Cy_0=y_1+y_{-1}.$$
See, for example, \cite{DDT}, \cite{E}, \cite{shin3}, \cite{shin1}.
Substituting $z=0$ we obtain
\begin{equation}\label{sd1}
C(\lambda)=\frac{\omega^{1/2}f(\omega^2\lambda)+
\omega^{-1/2}f(\omega^{-2}\lambda)}{f(\lambda)}.
\end{equation}
It is a very restrictive property of an entire function $f$, that
the zeros of denominator of (\ref{sd1}) are canceled by
the zeros of the numerator. In fact this property characterizes
our $f$ under mild additional conditions, see the discussion in \cite{E}.

\begin{figure}
\centering
\includegraphics[width=2.5in]{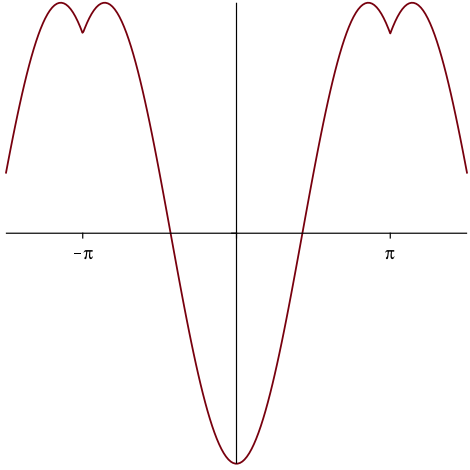}
\caption{Indicator of $f$ when $m\in(1,2)$. The jump of the slope occurs
at the accumulation point of the arguments of the zeros.
Here $m=1.5$.}\label{fig4}
\end{figure}

To investigate the zeros of $C$ we study the asymptotics of
the numerator. The sum in this numerator can be zero only near the
rays where asymptotics of the summands have equal absolute values.
It is useful to consider the
indicators of our entire functions \cite{L}.
In general, if $w$ is an entire
function of order $\rho$, normal type, the indicator is defined
by
$$h_w(\theta)=\limsup_{r\to\infty}r^{-\rho}\log|w(re^{i\theta})|,$$
so it shows the rate of growth of the function on the rays
$\arg\lambda=\theta$. For all our functions the limit in the
definition of the indicator exists for all rays except finitely many.
It follows from (\ref{3a}), (\ref{4a}) that
$$h_f(\theta)=K_m\cos\rho\theta,\quad |\theta|\leq\pi.$$
This indicator is shown in Fig.~\ref{fig4}. We recall that $K_m<0$ for
$m\in(1,2)$, the case considered in this section.
Indicators are always $2\pi$-periodic and continuous, and obey the following relations:
$$h_{w_1w_2}=h_{w_1}h_{w_2},$$
$$h_{w_1+w_2}\leq\max\{ h_{w_1},h_{w_2}\},$$
and equality holds in the last inequality for each $\theta$ for which
$h_{w_1}(\theta)\neq h_{w_2}(\theta)$.
In our simple case, all indicators are piecewise trigonometric:
they are solutions of the equation $h''+\rho^2h=0$ at every point
except finitely many. The finitely many points where $h'$ is discontinuous
are the accumulation points of the arguments of zeros.

\begin{figure}
\centering
\includegraphics[width=2.5in]{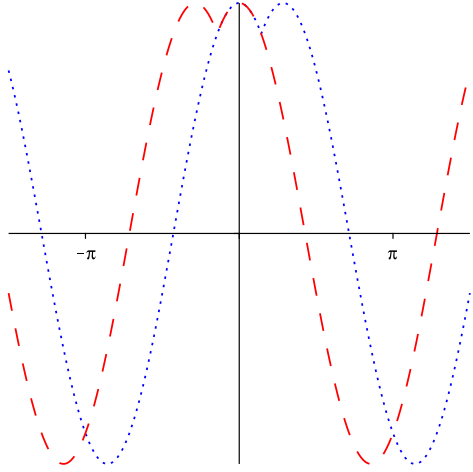}
\caption{Indicators of $\omega^{1/2}f(\omega^2\lambda)$ and
$\omega^{-1/2}f(\omega^{-2}\lambda)$.}\label{fig5}
\end{figure}

Indicators of the summands in the numerator of (\ref{sd1}) are
shown in Fig.~\ref{fig5}. The indicator of the sum will be equal to the
maximum of the indicators of the summands,
if we show that there is no cancellation on the interval
$$-\pi\frac{2-m}{2+m}<\theta<\pi\frac{2-m}{2+m},$$
where the indicators of the summands are equal.
To do this we use the asymptotics (\ref{3a}), (\ref{4a}).

So assume that
\begin{equation}\label{assumpt}
|\arg \lambda|<\pi\frac{2-m}{2+m},
\end{equation}
and recall that $\rho$ is defined by (\ref{rho}).
In computation with formula (\ref{4a}) one has to take into
account that the principal branch is used for the powers of $\lambda$,
so we have to replace every argument by a value on $(-\pi,\pi)$.
Thus, for example
\begin{equation}\label{omega2}
\arg(\omega^2\lambda)=\frac{4\pi}{m+2}+\arg\lambda-2\pi=-\frac{2\pi m}{m+2}+\arg\lambda\in (-\pi,\pi).
\end{equation}
Now we have using (\ref{omega2}):
$$\omega^{1/2}(\omega^2\lambda)^{-1/4}=i\lambda^{-1/4},$$
and
$$\omega^{-1/2}(\omega^{-2}\lambda)^{-1/4}=-i\lambda^{-1/4}.$$
Similarly
$$(\omega^2\lambda)^\rho=(\omega^{-2}\lambda)^\rho=-\lambda^\rho,$$
and more generally, for integer $k$:
$$(\omega^2\lambda)^{-k\rho}=(\omega^{-2}\lambda)^{-k\rho}=
(-1)^k\lambda^{-k\rho},$$
In particular, this shows that all terms $\lambda^{k\rho}$
with integer $k$ cancel in the asymptotics of the sum
$\omega^{1/2}f(\omega^2\lambda)+\omega^{-1/2}f(\omega^{-2}\lambda).$
However
$$(\omega^{\pm 2}\lambda)^{-m\rho}=\omega^{\mp 2m\rho}\lambda^{-m\rho}=
e^{\pm i\pi m}\lambda^{-m\rho},$$
where we used (\ref{omega2}).
Combining all these calculations, we conclude that
$$\omega^{1/2}f(\omega^2\lambda)+\omega^{-1/2}f(\omega^{-2}\lambda)=
-2\exp(-K_m\lambda^\rho)(\lambda^{-m\rho}\sin\pi m +o(\lambda^{-m\rho})),$$
when $\lambda$ satisfies (\ref{assumpt}).
This shows that when $m$ is not an integer, there is no complete cancellation
of the asymptotic expansions, and the indicator of
the numerator in (\ref{sd1}) is the maximum of the indicators
of the summands, and also shows that the spectral determinant $C(\lambda)$
has no zeros in the sector (\ref{assumpt}). All this is true when $m\in(1,2)$.
\vspace{.1in}

\begin{figure}
\centering
\includegraphics[width=2.5in]{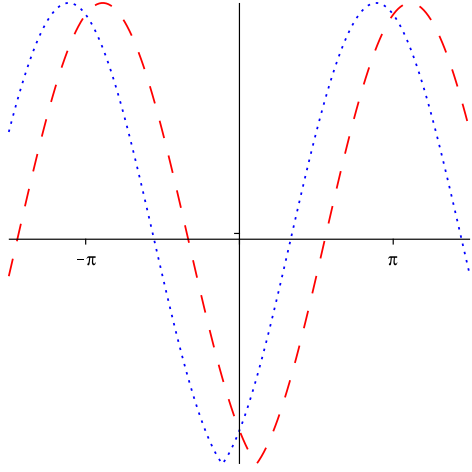}
\caption{Indicators of $\omega^{1/2}f(\omega^2\lambda)$
and $\omega^{-1/2}f(\omega^{-2}\lambda)$ for $m>2$. Here $m=2.5$.}
\label{fig6}
\end{figure}

\begin{figure}
\centering
\includegraphics[width=2.5in]{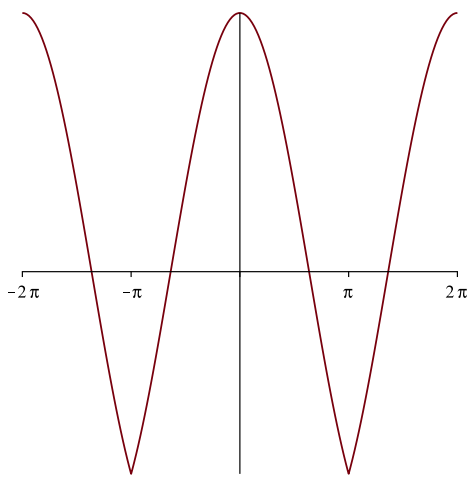}
\caption{Indicator of $f$ for $m>2$. Here $m=3.5$.}\label{fig7}
\end{figure}

{\em Remark 1.}
To see what is the difference between the cases $m<2$ and  $m>2$,
we sketch the indicators of
the numerator in this case in Fig.~\ref{fig6}; notice that for $m>2$,
the indicator of $f$ looks like Fig.~\ref{fig7}, because $K_m>0$.
Fig.~\ref{fig6} shows that
there are zeros of the numerator of (\ref{sd1}) whose arguments
accumulate to $0$, and in fact there are infinitely many positive
zeros as proved in \cite{E}.
\vspace{.1in}

Now we return to the proof of Theorem 1.
Indicator of the numerator in (\ref{sd1}) in Fig.~\ref{fig5} shows that the only
possible direction of accumulation of the arguments of
zeros of $C$ can be the negative ray $\arg\lambda=\pi$, and the
rays $\arg\lambda=\pm\pi(2-m)/(m+2)$.
Indeed there are
infinitely many zeros of the numerator with arguments accumulating
to these three points. We will show that zeros with arguments
near $\pi$ are actually real,
and {\em almost all of them cancel} with the zeros of the denominator $f$.

To do this we first show that the asymptotics of the numerator and
denominator of (\ref{sd1}) differ by an exponential factor.
Indeed, for
\begin{equation}\label{s2}
\pi-\epsilon<\arg z<\pi+\epsilon,
\end{equation}
where $\epsilon$ is sufficiently small, we have
in view of (\ref{shneg})
\begin{equation}\label{shnegp}
f(\lambda)=\lambda^{-1/4}\exp\left( K_m\lambda^\rho+O(\lambda^{-\rho})\right)+
i\lambda^{-1/4}
\exp\left(-K_me^{-2\pi i/m}\lambda^\rho+O(\lambda^{-\rho})\right),
\end{equation}
where we used the fact that $e^{-2\pi i\rho}=-e^{-2\pi i/m}.$
On the other hand, in the same sector (\ref{s2}) we have
\begin{eqnarray*}
\omega^{1/2}f(\omega^2\lambda)+\omega^{-1/2}f(\omega^{-2}\lambda)
&=&i\lambda^{-1/4}\exp\left(-K_m\lambda^\rho+O(\lambda^{-\rho})\right)\\
&+&\lambda^{-1/4}\exp\left( K_me^{-2\pi i/m}\lambda^\rho+O(\lambda^{-\rho})\right),
\end{eqnarray*}
and multiplying this by the exponential factor
$$\exp\left(((1-e^{-2\pi i/m})K_m\lambda^\rho\right),$$
we obtain (\ref{shnegp}).

Now we follow the arguments from \cite{shin1}.
Let $\lambda$ be a root of $f$.
We conclude from (\ref{shneg}), (\ref{shneg1})
(which is the same as (\ref{shnegp})) that
\begin{equation}\label{equ}
\exp\left( K_m\lambda^\rho+O(\lambda^{-\rho})\right)=
-i\exp\left( K_me^{-2\pi i\rho}+O(\lambda^{\rho})\right).
\end{equation}
Denoting $\lambda^\rho=e^{\pi i\rho}\zeta$, we obtain
$$\zeta_n=an+b+O(n^{-1}),$$
where
$$a=\frac{\pi}{K_m\sin\pi\rho},\quad b=-\frac{\pi}{4K_m\sin\pi\rho}.$$
This implies the approximate formula for the roots of (\ref{equ})
\begin{equation}\label{asr}
-\lambda_n=(an)^{1/\rho}+(b/\rho)(an)^{1/\rho-1}+O(n^{1/\rho-2}).
\end{equation}
As the asymptotics of $f$ and of the numerator of (\ref{sd1})
differ only by an exponential multiple, we conclude that the same
formula
(\ref{asr}), with the same $a$ and $b$,
also holds for the roots of the numerator of (\ref{sd1}) near the negative ray.
The error term $O(n^{1/\rho-2})$ in this formula tends to zero
faster than
the distance between the neighboring roots. Indeed this distance is
at least
$$c_1\left((n+1)^{1/\rho}-n^{1/\rho}\right)>c_2n^{1/\rho-1},$$
where $c_1$ and $c_2$ are some positive constants.
As $C$ is entire, each root of $f$ must cancel with some root of
the numerator of (\ref{sd1}), and we conclude that {\em almost all}
roots of this numerator near the negative ray cancel.
Therefore, $C$ has finitely many
roots near the negative ray.

This completes the proof of Theorem 1.
\vspace{.1in}

\noindent
{\bf Problem of level 2. Proof of Theorem 2.}
\vspace{.1in}

\begin{figure}
\centering
\includegraphics[width=2.5in]{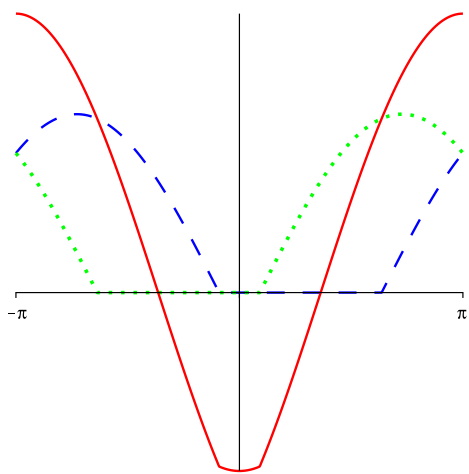}
\caption{Indicators of the three summands in the numerator of
(\ref{D}).}\label{fig8}
\end{figure}

The spectral determinant of the level $2$ problem is
\begin{equation}\label{D}
D(\lambda)=\frac{\omega f(\omega\lambda)f(\omega^3\lambda)+
\omega^{-1}f(\omega^{-1}\lambda)f(\omega^{-3}\lambda)+
f(\omega^3\lambda)f(\omega^{-3}\lambda)}{f(\omega\lambda)f(\omega^{-1}\lambda)},
\end{equation}
see, for example \cite{E}. This is obtained from the formula
$$D(\lambda)=C(\omega\lambda)C(\omega^{-1}\lambda)-1,$$
which is derived in \cite{DDT}, \cite{E} and \cite{shin3},
by substituting the expression (\ref{sd1}) for $C$.
The order $\rho$ of $f$ is less than $1$ now,
and the indicator of
$f$ is shown in Fig.~\ref{fig7}.
The indicators of the three summands in the numerator on $D$ are shown
in Fig.~\ref{fig8}.

We have to verify that there is no cancellation, so that the indicator
of the numerator in (\ref{D}) is indeed zero on the interval
around $0$. We compute the asymptotic expansion of this numerator,
neglecting the third summand $f(\omega^3\lambda)f(\omega^{-3}\lambda)$
whose indicator is negative on the
interval under consideration.

Beginning with $\omega f(\omega\lambda)f(\omega^3\lambda)$, we use (\ref{4a}),
and restrict $\lambda$ to the sector
\begin{equation}\label{sector3}
|\arg \lambda|<\pi\frac{4-m}{2+m}.
\end{equation}
For such $\lambda$, we have the following principal values
of the arguments:
$$\arg\omega\lambda=\frac{2\pi}{m+2}+\arg\lambda,$$
$$\arg\omega^3\lambda=\frac{6\pi}{m+2}-2\pi+\arg\lambda=-\frac{2\pi(m-1)}{m+2}+\arg\lambda\in(-\pi,\pi).$$
Using (\ref{rho}), we conclude that
$$\omega^\rho+(\omega^3)^\rho=0,$$
therefore the exponential term in
$\omega f(\omega\lambda) f(\omega^3\lambda)$ disappears,
as seen in Fig.~\ref{fig8}.
The multiple coming from $\lambda^{-1/4}$ is
$$\omega(\omega\lambda)^{-1/4}(\omega^3\lambda)^{-1/4}=
i\lambda^{-1/2},$$
and similarly
$$\omega^{-1}(\omega^{-1}\lambda)^{-1/4}(\omega^{-3}\lambda)^{-1/4}=-i\lambda^{-1/2}.$$
The rest is an asymptotic series in negative powers of $\lambda^\rho$.
We claim that the top non-integer power of $\lambda^{\rho}$ in
this series for the numerator of (\ref{D}) does not vanish.
Indeed, this term is $\Gamma(m+1)/8$ multiplied by
\begin{eqnarray*}
&&i\left((\omega\lambda)^{-m\rho}+(\omega^3\lambda)^{-m\rho}\right)
-i\left((\omega^{-1}\lambda)^{-m\rho}+(\omega^{-3}\lambda)^{-m\rho}\right)\\
&=&i\left((e^{i\pi m}-e^{-i\pi m}\right)\lambda^{-m\rho}=
2\lambda^{-m\rho}\sin\pi m\neq 0,
\end{eqnarray*}
when $m$ is not an integer.

Thus $D$ has infinitely many zeros with arguments accumulating
to the boundary rays of the sector (\ref{sector3}). The numerator also has zeros with
arguments accumulating to $\pm 2\pi/(m+2)$ but almost all these zeros
cancel with the zeros of the numerator. This is proved by a reasoning
similar to that in the previous section.
This completes the proof of our theorem.
\vspace{.1in}

\noindent
{\em Remark 2.} When $m$ is an integer, the asymptotic expansion in (\ref{4a})
contains only integer powers, and we {\em must} have complete cancellation
of all terms of the asymptotic expansion of the numerator of (\ref{D}).
\vspace{.1in}

\noindent
{\em Remark 3.} It is very plausible that the pattern proved in this paper
persists for all levels, and the pictures of indicators of the spectral
determinants suggest that the same proof can be extended.
\vspace{.1in}

The authors thank Victor Katsnelson,
Fedja Nazarov and Kwang Shin for useful discussions.

{\em Department of Mathematics

Purdue University

West Lafayette, IN 47907

USA

eremenko@math.purdue.edu

agabriel@math.purdue.edu}

\begin{thebibliography}{1}

\bibitem{BB0} C. Bender and S. Boettcher, Real spectra
in non-Hermitian Hamiltonians having PT symmetry. Phys. Rev. Lett. 80 (1998),
no. 24, 5243--5246.
\bibitem{BB} C. Bender and S. Boettcher, PT-symmetric quantum mechanics,
J. Math. Phys. 40 (1999), no. 5, 2201--2229.
\bibitem{Bn} C. Bender, N. Hassanpour, D. Hook, S. Klevansky, C. S\"underhauf,
and Zichao Wen, Behavior of eigenvalues in a region with broken
PT symmetry, Phys. Rev. A 95 (2017) 052113.
\bibitem{BW} C. Bender and Tai Tsun Wu, Tsun Anharmonic oscillator,
Phys. Rev. (2) 184 1969 1231--1260.
\bibitem{BJ} M. Born and P. Jordan, Zur Quantummechenik,
Z. Phys., 34 (1925) p. 858.
English transl.:
V. L. van der Waerden, Sources of quantum mechanics,
North Holland, Amsterdam, 1967, 277--306.
\bibitem{Bourbaki} N. Bourbaki, Fonctions d'une variable r\'eelle,
Herman, Paris, 1951.
\bibitem{DDT0} P. Dorey, C. Dunning and R. Tateo,
Spectral equivalences, Bethe ansatz equations, and reality properties in
PT-symmetric quantum mechanics. J. Phys. A 34 (2001), no. 28, 5679--5704.
\bibitem{DDT} P. Dorey, C. Dunning and R. Tateo,
The ODE/IM correspondence. J. Phys. A 40 (2007), no. 32, R205--R283.
\bibitem{DDT3} P. Dorey, C. Dunning and R. Tateo,
From PT-symmetric quantum mechanics to conformal field theory,
Pramana -- journal of physics, 73 (2009) 217--239.
\bibitem{DDT2} P. Dorey,  A. Millican-Slater and R. Tateo,
Beyond the WKB approximation in PT-symmetric quantum mechanics,
J. Phys. A 38 (2005), no. 6, 1305--1331.
\bibitem{E} A. Eremenko, Entire functions, PT-symmetry
and Voros's quantization scheme, arXiv:1510.02504.
\bibitem{Heading} J. Heading, Introduction to phase-integral methods,
John Willey \&\ Sons, Inc., NY 1962.
\bibitem{Fed} M. V. Fedoryuk, Asymptotics. Integrals and series (Russian)
Moscow, Nauka, 1987.
\bibitem{Heis} W. Heisenberg, \"Uber quantentheoretische Umdeutung
kinematischer und mechanischer Bezeichingen, Z. Phys., 33 (1925)
p. 839. English transl.:
V. L. van der Waerden, Sources of quantum mechanics,
North Holland, Amsterdam, 1967, 261--276.
\bibitem{L} B. Ya Levin, Distribution of zeros of entire functions,
AMS, Providence RI, 1980.
\bibitem{shin3} K. Shin, The potential $(iz)^m$ generates real
eigenvalues only under symmetric rapid decay conditions,
J. Math. Phys., 46 (2005) 082110.
\bibitem{shin1} K. Shin, Asymptotics of eigenvalues of non-self-adjoint
Schr\"odinger operators on a half-line, Comp. methods and function theory,
10 (2010) 2, 111--133.
\bibitem{Sibuya} Y. Sibuya, Global theory of a second order linear
ordinary differential equation with a polynomial coefficient,
North-Holland, NY, 1975.
\bibitem{Tabara} T. Tabara, Asymptotic behavior of Stokes multipliers for
$y''-(x\sigma+\lambda)y=0\; (\sigma\geq 2)$ as $\lambda\to\infty$,
Differential equations and dynamical systems (Waterloo, ON, 1997).
Dynam. Contin. Discrete Impuls. Systems 5 (1999), no. 1-4, 93--105.
\bibitem{Voros} A. Voros, The return of the quartic oscillator:
the complex WKB method,
Ann. Inst. H. Poincar\'{e} Sect. A, 39 (1983), no. 3, 211--338.
\end{thebibliography}
\end{document}